\documentclass[aps,showpacs]{revtex4}

\usepackage[dvips]{graphics}

\begin{document}


\def\beq{\begin{equation}}
\def\eeq{\end{equation}}
\def\bea{\begin{eqnarray}}
\def\eea{\end{eqnarray}}
\def\ket#1{|#1\rangle}
\def\bra#1{\langle#1|}


\title{\Large\bf Distributing a multiparticle state
                 by entanglement swapping}

\author{{\bf Chi-Yee Cheung}}
\email{cheung@phys.sinica.edu.tw}

\affiliation{Institute of Physics, Academia Sinica\\
             Taipei 11529, Taiwan, Republic of China\\}


\begin{abstract}

Using entanglement swapping, we construct a scheme to
distribute an arbitrary multiparticle state to remote
receivers. Only Bell states and two-qubit collective
measurements are required.

\end{abstract}


\maketitle


The superposition principle is the most novel and least
understood feature of the quantum theory. Quantum
entanglement arises from the superposition of multiparticle
states. For particles located far apart from one another,
quantum entanglement gives rise to the mysterious
phenomenon of non-local correlation to which there exists
no local classical explanation \cite{EPR,Bell,CHSH,Aspect}.
With the advent of quantum information science in recent
years, instead of remaining merely a mystery to be solved,
quantum entanglement has become a valuable and often
indispensable resource in every branch of the field.

A bipartite state $\ket{\psi_{1,2}}$ is entangled if it
cannot be written in a product form:
 \beq
 \ket{\psi_{1,2}}=\ket{\psi'_1}\otimes\ket{\psi''_2}.
 \eeq
The two-qubit Bell (or EPR) states,
 \bea
 &&\ket{\phi^{\pm}_{1,2}}={1\over\sqrt{2}}
 \Big(\ket{0_1} \ket{1_2}\pm\ket{0_1} \ket{1_2}\Big),\\
 &&\ket{\varphi^{\pm}_{1,2}}={1\over\sqrt{2}}
 \Big(\ket{0_1} \ket{0_2}\pm\ket{1_1} \ket{1_2}\Big),
 \eea
are the most common entangled states employed in quantum
information science. (The $\otimes$ sign will be understood
hereafter.) Usually two separated particles are entangled
because they were once in contact and interacted with each
other. However, by entanglement swapping
\cite{BVK,ZZHE,JWPZ}, we can entangle two remotely
separated particles which do not have a history of mutual
interaction. This is a vivid demonstration of nonlocal
correlations in the quantum theory. The idea of
entanglement swapping is mathematically very simple.
Consider two Bell states, say $\ket{\varphi^+_{1,2}}$ and
$\ket{\phi^-_{3,4}}$, where particles 1 and 2 have never
been in contact with particles 3 and 4 before. The product
of these two states can be rewritten as
 \beq
 \ket{\varphi^+_{1,2}}\ket{\phi^-_{3,4}}
 =\ket{\varphi^+_{1,3}}\ket{\phi^-_{2,4}}
 +\ket{\varphi^-_{1,3}}\ket{\phi^+_{2,4}}
 -\ket{\phi^+_{1,3}}\ket{\varphi^-_{2,4}}
 -\ket{\phi^-_{1,3}}\ket{\varphi^+_{2,4}}.
 \label{product}
 \eeq
Therefore if we make a Bell measurement on the pair (1,3),
then depending on the outcome the (2,4) pair would collapse
into one of the corresponding Bell states:
$\ket{\phi^-_{2,4}}$, $\ket{\phi^+_{2,4}}$,
$\ket{\varphi^-_{2,4}}$, and $\ket{\varphi^+_{2,4}}$. In
other words, particles 2 and 4 would become entangled even
though they are far apart and have never interacted with
each other in the past.

In this paper, we shall use entanglement swapping to
distribute an arbitrary $N$-particle state to $M\le N$
remote parties. Distribution of quantum information is
essential in quantum secret sharing (QSS) and other
applications in quantum information science, such as
quantum communication network and distributed quantum
computation \cite{CEHM}. QSS is the quantum counterpart of
classical secret sharing first discussed by Blakely
\cite{Blakely} and Shamir \cite{Shamir} in 1979. The idea
is as follows. Suppose Alice wants to send a secret message
to a remote location, and she can send it to either of her
two agents, Bob and Charlie. As a precaution against
information leakage and misuse, it is safer for her to
split the message into two pieces and send them separately
to Bob and Charlie, such that anyone alone has absolutely
no knowledge of the message. Bob and Charlie can
reconstruct the original secret message only if they
collaborate with each other. Clearly the above
consideration can be generalized to secret sharing by $N$
parties.

QSS refers to the implementation of the secret sharing task
outlined above using quantum mechanical resources. Hillary
{\it et al.} \cite{HBB} and Karlsson {\it et al.}
\cite{KKI} were the first to propose QSS protocols using
respectively three-particle Greenberger-Horne-Zeilinger
(GHZ) states and two-particle Bell states. Since then, a
wide variety of other QSS protocols have been proposed.
Although the goal of most QSS protocols is to protect a
classical secret message, the notion of QSS has also been
generalized to the sharing of a secret quantum state
\cite{HBB,KKI,CGL,LSBSL}, which is also referred to as
``quantum state sharing".

Whether the goal is to share a classical or a quantum
secret, in almost all cases, it is necessary to distribute
a $N$-qubit entangled state among $M$ parties, where $N\ge
M\ge 2$. For example, in the QSS protocol of Hillery {\it
et al.} \cite{HBB}, Alice uses GHZ states to split a
quantum key into two shares such that each of the two
agents, Bob and Charlie, gets only one share. To do so,
Alice must be able to safely distribute two of the
particles in each tripartite GHZ state separately to Bob
and Charlie. In the quantum state sharing protocol proposed
by Cleve {\it et al.} \cite{CGL}, the secret message to be
shared is a single qutrit state. The protocol is similar to
an error-correcting code, in which a three-qutrit entangled
state is generated from the secret qutrit. The ``dealer"
then distributes the resulting three qutrits to three
different parties. More generally, in a so-called $(k,n)$
threshold scheme, the secret (quantum or classical) is
divided into $n$ shares, such that any $k$ of those shares
can be used to reconstruct the secret, while any set of
less than $k$ shares contains absolutely no information
about the secret at all. So in general one needs to
distribute an arbitrary entangled state to $n$ different
parties.

Of course Alice could send the particles involved directly
over quantum channels to their respective destinations.
This practice is however both inefficient and unsafe. First
of all, since noise is always present in any available
channel, quantum information may get distorted; more
seriously decoherence effects may even cause the particle
to collapse. The loss of any one particle in an unknown
multiparticle state due to decoherence or dissipation will
require the whole state to be regenerated again. Hence
direct distribution of the particles is not an efficient
way to proceed. Furthermore the state to be shared is
itself the carrier of information, therefore for security
reasons, it is not safe to send them directly over
long-distance quantum channels which may be monitored by
eavesdroppers. Of course eavesdropping activities can be
detected by security testing methods, but they inevitably
involve measuring some of the particles going through the
channels, which is obviously not acceptable if they are
part of the multiparticle state being distributed.

It is therefore more desirable to distribute the particles
using quantum entanglement plus local operations and
classical communications only. Entanglement resources are
usually supplied by two-qubit Bell states or sometimes
three-qubit GHZ states shared between the sender and the
intended receivers. The advantage of this approach is that
the required entanglement can be established and tested
independent of the state to be distributed. Once it is
securely established, quantum channels are no longer
needed, and transmission noise is no longer a problem.

In the following, we show how to faithfully distribute an
arbitrary $N$-qubit state using entanglement swapping; our
scheme generalizes a multipartite QSS protocol discussed in
Ref. \cite{KKI}. Schemes of distributing $N$-qubit states
by teleportation have also been proposed
\cite{Rigolin,IZZ}. However, these protocols have the
undesirable features that the complexity of the required
collective measurements increases with $N$. In our scheme,
only Bell states are used and only two-qubit collective
measurements are required.

To proceed, we first show that entanglement swapping
between two Bell states can be generalized to that between
an arbitrary $N$-qubit state and a Bell state. The setting
is that Alice owns an arbitrary $N$-qubit state
$\ket{\Psi_{1,...,N}}$, and she shares a Bell state
$\ket{\phi^-_{\mu,\nu}}$ with Bob who is somewhere far
away; Alice holds qubit-$\mu$ and Bob qubit-$\nu$. From
 \beq
 \ket{\Psi_{1,...,N}}=
 \big(\ket{0_i}\bra{0_i}+\ket{1_i}\bra{1_i}\big)
 \ket{\Psi_{1,...,N}},\qquad (1\le i\le N),
 \eeq
we see that $\ket{\Psi_{1,...,N}}$ can always be rewritten
as
 \beq
 \ket{\Psi_{1,...,N}}
 =a\ket{0_i}\ket{\Phi_{1,...,i-1,i+1,...,N}}
 +b\ket{1_i}\ket{\Phi'_{1,...,i-1,i+1,...,N}},
 \label{Psi}
 \eeq
where $|a|^2+|b|^2=1$, and
$\ket{\Phi_{1,...,i-1,i+1,...,N}}$ and
$\ket{\Phi'_{1,...,i-1,i+1,...,N}}$ are normalized states
of $(N-1)$ qubits. Note that unless $a$ or $b$ vanishes, or
$\ket{\Phi_{1,...,i-1,i+1,...,N}}$ and
$\ket{\Phi'_{1,...,i-1,i+1,...,N}}$ differ only by a phase
factor, otherwise qubit-$i$ is entangled with the rest of
the group.

Similar to Eq. (\ref{product}), we can rewrite the product
of $\ket{\Psi_{1,...,N}}$ and $\ket{\phi^-_{\mu,\nu}}$ as
 \bea
 \ket{\Psi_{1,...,N}}\ket{\phi^-_{\mu,\nu}}
 ={1\over 2}&\Big[&
 \ket{\varphi^+_{i,\mu}}
 \Big(a\ket{1_\nu}\ket{\Phi_{1,...,i-1,i+1,...,N}}
 -b\ket{0_\nu}\ket{\Phi'_{1,...,i-1,i+1,...,N}}\Big)
 \nonumber\\*
 &+&\ket{\varphi^-_{i,\mu}}
 \Big(a\ket{1_\nu}\ket{\Phi_{1,...,i-1,i+1,...,N}}
 +b\ket{0_\nu}\ket{\Phi'_{1,...,i-1,i+1,...,N}}\Big)
 \nonumber\\*
 &-&\ket{\phi^+_{i,\mu}}
 \Big(a\ket{0_\nu}\ket{\Phi_{1,...,i-1,i+1,...,N}}
 -b\ket{1_\nu}\ket{\Phi'_{1,...,i-1,i+1,...,N}}\Big)
 \nonumber\\*
 &-&\ket{\phi^-_{i,\mu}}
 \Big(a\ket{0_\nu}\ket{\Phi_{1,...,i-1,i+1,...,N}}
 +b\ket{1_\nu}\ket{\Phi'_{1,...,i-1,i+1,...,N}}\Big)\Big].
 \eea
Hence a Bell measurement by Alice on the $(i,\mu)$ pair
will entangle the remote qubit-$\nu$ to the local group of
($N-1$) qubits $(1,...,i-1,i+1,...,N)$. This process is
depicted diagrammatically in Fig. 1.

\begin{figure}
\begin{center}
\scalebox{0.5}{\includegraphics{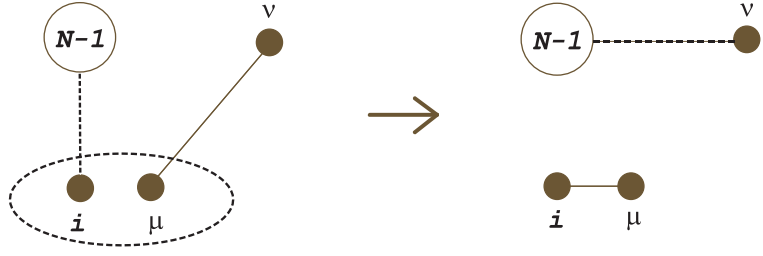}}
\caption{Entanglement swapping between an arbitrary
$N$-qubit state and a Bell state ($\mu,\nu$). Qubit-$\nu$
belongs to Bob, and the rest to Alice; dashed (solid)
straight lines indicate (maximal) entanglement, and dashed
ellipse represents a Bell measurement.}
\end{center}
\end{figure}

The resulting $N$-qubit state depends on the outcome of the
Bell measurement:
 \bea
 &&(1)\quad\ket{\varphi^+_{i,\mu}}~\longrightarrow~
 a\ket{1_\nu}\ket{\Phi_{1,...,i-1,i+1,...,N}}
 -b\ket{0_\nu}\ket{\Phi'_{1,...,i-1,i+1,...,N}},
 \label{outcome-1}\\
 &&(2)\quad\ket{\varphi^-_{i,\mu}}~\longrightarrow~
 a\ket{1_\nu}\ket{\Phi_{1,...,i-1,i+1,...,N}}
 +b\ket{0_\nu}\ket{\Phi'_{1,...,i-1,i+1,...,N}},
 \label{outcome-2}\\
 &&(3)\quad\ket{\phi^+_{i,\mu}}~\longrightarrow~
 a\ket{0_\nu}\ket{\Phi_{1,...,i-1,i+1,...,N}}
 -b\ket{1_\nu}\ket{\Phi'_{1,...,i-1,i+1,...,N}},
 \label{outcome-3}\\
 &&(4)\quad\ket{\phi^-_{i,\mu}}~\longrightarrow~
 a\ket{0_\nu}\ket{\Phi_{1,...,i-1,i+1,...,N}}
 +b\ket{1_\nu}\ket{\Phi'_{1,...,i-1,i+1,...,N}}
 \label{outcome-4}.
 \eea
Comparing with Eqs. (\ref{Psi}), we see that if Alice tells
Bob which of the four Bell states she has obtained, then by
a local unitary transformation on qubit-$\nu$ Bob can
rotate the state of the $N$-qubit group
$(1,...,i-1,\nu,i+1,...,N)$ back to the original state
$\ket{\Psi_{1,...,N}}$ (with qubit-$i$ $\rightarrow$
qubit-$\nu$). The required unitary operators are
($\sigma_z\sigma_x$, $\sigma_x$, $\sigma_z$, $I$)
respectively for the four possible outcomes listed in Eqs.
(\ref{outcome-1}$-$\ref{outcome-4}). It is easily seen that
the information cost for the whole operation is one e-bit
plus two c-bits (classical bits) from Alice to Bob. If
qubit-$i$ is not entangled with the group
$(1,...,i-1,i+1,...,N)$, that is, if Eq. (\ref{Psi}) can be
reduced to
 \beq
 \ket{\Psi_{1,...,N}}
 =\big(a\ket{0_i}+b\ket{1_i}\big)
 \ket{\Phi_{1,...,i-1,i+1,...,N}},
 \eeq
then what we have done is simply the teleportation
\cite{BBCJPW} of a single qubit $\big(a\ket{0_i}
+b\ket{1_i}\big)$ from Alice to Bob. For a general
$\ket{\Psi_{1,...,N}}$ our procedure effectively teleports
the entanglement to Bob.

Notice that the above procedure is entirely general, in the
sense that it is independent of the state of the inactive
qubits $(1,...,i-1,i+1,...,N)$. Therefore it can be
repeated until all the qubits are distributed to their
respective remote locations as desired. We note in passing
that, if Alice leaves some of the qubits undistributed,
then the distributed qubits form a mixed state.

It is interesting to note that each round of the operation
(Bell measurement plus unitary rotation) described above
can be summarized by the action of an operator $U_{i\nu}$
which interchanges the states of qubit-$i$ and qubit-$\nu$:
 \beq
 U_{i\nu}\big( \ket{\Psi_{1,...,i-1,i,i+1,...,N}}
 \ket{\phi^-_{\mu,\nu}}\big) ~=~
 \ket{\Psi_{1,...,i-1,\nu,i+1,...,N}}
 \ket{\phi^-_{\mu,i}},
 \label{swap}
 \eeq
where without loss of generality we have assumed that the
measured Bell state has been rotated back to the singlet
Bell state $\ket{\phi^-_{\mu,i}}$, which Alice can easily
do by a local unitary transformation. The effect of
$U_{i\nu}$ is identical to that of a regular two-qubit swap
gate \cite{NC} except for the fact that here the qubits
involved are remotely separated; hence $U_{i\nu}$ is a kind
of nonlocal swap operator. The whole distribution process
can be accomplished by simply repeating this swapping
operation $N$ times:
 \beq
 \prod_{i=1}^N U_{i\nu_i}\Big( \ket{\Psi_{1,...,N}}
 \prod_{i=1}^N\ket{\phi^-_{\mu_i,\nu_i}}\Big) ~=~
 \ket{\Psi_{\nu_1,...,\nu_N}}
 \prod_{i=1}^N\ket{\phi^-_{\mu_i,i}},
 \eeq
where the $N$ qubits $(\nu_1,...,\nu_N)$ are located at
their respective destinations.

From this perspective, it is clear that we could also use
the general nonlocal swap operation \cite{CLP,EJPP} in
place of $U_{i\nu_i}$, although it would be rather
inefficient to do so. It is known that the implementation
of a general nonlocal swap operation requires at least two
e-bits plus two c-bits from Alice to Bob plus another two
c-bits from Bob to Alice \cite{CLP,EJPP}. However in our
scheme, only one e-bit and two c-bits from Alice to Bob are
needed per swap. This difference in resources consumption
is due to the fact that $U_{i\nu_i}$ is actually not an
universal swap operator, because it cannot be used to swap
two arbitrary remote qubits. Our scheme is only good for
the special purpose of qubit distribution, where each round
of swapping is analogous to the teleportation of a single
qubit, hence relatively less resources are required.

In summary, we have constructed a scheme to distribute an
arbitrary $N$-qubit state (pure or mixed) to $M\le N$
remote parties. The basic operation used is entanglement
swapping which involves only Bell states and two-qubit
collective measurements. Our scheme is experimentally
feasible with currently available technologies
\cite{LSBSL,JWPZ}.





\end{document}